\documentclass[a4paper]{jpconf}
\usepackage{graphicx}
\usepackage[utf8x]{inputenc}
\usepackage{amsmath}
\usepackage{SIunits}
\newcommand{\bra}[1]{\langle #1 \hspace{-2pt} \mid}
\newcommand{\ket}[1]{\mid \hspace{-1pt} #1 \rangle}
\newcommand{\abs}[1]{\ensuremath{\left\vert #1 \right\vert}}
\newcommand{\D}{\mathrm{d}}
\newcommand{\erw}[1]{\langle #1 \rangle}
\renewcommand{\vec}[1]{\mathbf{#1}}
\begin{document}
\title{Newtonian self-gravity in trapped quantum systems and experimental tests}

\author{Andr\'e Gro{\ss}ardt}

\address{Zentrum f\"ur angewandte Raumfahrttechnologie und Mikrogravitation (ZARM),\\
Universit\"at Bremen, Am Fallturm 2, 28359 Bremen, Germany}
\address{Dipartimento di Fisica, Universit\`a degli Studi di Trieste,\\
Strada Costiera 11, 34151 Miramare-Trieste, Italy}

\ead{andre.grossardt@zarm.uni-bremen.de}

\begin{abstract}
No experimental evidence exists, to date, whether or not the gravitational field must be quantised.
Theoretical arguments in favour of quantisation are inconclusive. The most straightforward
alternative to quantum gravity, a coupling between classical gravity and quantum matter according
to the semi-classical Einstein equations, yields a nonlinear modification of the Schrödinger
equation. Here, effects of this so-called Schrödinger-Newton equation are discussed, which allow
for technologically feasible experimental tests.
\end{abstract}

%% 8 PAGES TOTAL!

\section{Introduction}
The question how to reconcile the foundations of quantum theory with the dynamical space-time structure
of general relativity is often reduced to problems arising at the Planck scale, e.\,g. in the physics
of black holes or the early universe. Nonetheless, the status of the description of both gravity
and quantum matter in a common framework is far from clear also at low energies: certainly from
the point of view of experimental observations, but also from a theoretical perspective.

An illustrative example is the familiar double-slit experiment, if conducted with a
massive\footnote{what quantifies a massive particle depends, of course, on the precise experimental
context, specifically on feasible force sensitivities} particle. Assume a high sensitivity
gravitational probe (a test mass attached to a sophisticated force metre) could be placed behind the
slit, sensing the gravitational field of the spatial superposition state prepared by the slit.
What is the gravitational field it would observe?

In analogy to electrodynamics one is tempted to think that the answer is obvious:
The gravitational field should be in a superposition state itself, which is entangled with the particle
state. However, the analogy between gravity and electrodynamics must eventually fail; at the latest
when it comes to the question of renormalisability. One may, therefore, ask whether this analogy is
still appropriate for the case of macroscopic non-classical states at low energy.
In general relativity, on the other hand, the gravitational field is an expression of the
space-time structure. Matter acts as a source, determining the curvature of space time.
In the presence of quantised matter fields, one would ask: how does this quantum matter
in a non-localised state source the gravitational field?

Keeping the classical space-time structure of general relativity, and treating quantum
fields---which are then living on this curved space-time---as a curvature source, gives rise to
the idea of (fundamentally) semi-classical gravity. The seemingly most straightforward way
to incorporate quantum matter as a source into Einstein's equations is to replace the
classical stress-energy tensor by the expectation value of the corresponding quantum operator:
\begin{equation}\label{eqn:sce}
R_{\mu \nu} - \frac{1}{2} \, g_{\mu \nu} \, R
= \frac{8 \pi \, G}{c^4} \, \langle \Psi \vert \hat{T}_{\mu \nu} \vert \Psi \rangle
\end{equation}
These are the semi-classical Einstein equations which have first been proposed in this context
by M\o{}ller~\cite{Moller:1962} and Rosenfeld~\cite{Rosenfeld:1963}.

Arguments have been invoked attempting to rule out this type of coupling between quantum matter
and gravity solely on theoretical grounds. It has been pointed out that Eq.~\eqref{eqn:sce}
is incompatible with both an instantaneous wave-function collapse and a no-collapse interpretation
of quantum mechanics~\cite{Page:1981}. Nonetheless, this leaves open the possibility of a dynamical
collapse in the spirit of collapse models~\cite{Bassi:2013}.
More general assertions against any type of semi-classical coupling~\cite{Eppley:1977}
were found inconclusive~\cite{Mattingly:2005,Kiefer:2007,Albers:2008}.
It seems that Rosenfeld was right when emphasising the necessity of experimental tests
whether or not the gravitational field should be quantised:
\emph{``There is no denying that, considering the universality of the quantum of action, it is very tempting
to regard any classical theory as a limiting case to some quantal theory. In the absence of empirical evidence,
however, this temptation should be resisted. The case for quantizing gravitation, in particular, far from being
straightforward, appears very dubious on closer examination.''}~\cite{Rosenfeld:1963}

It was Carlip~\cite{Carlip:2008} who first pointed out that an experimental test of semi-classical
gravity could be feasible in the nonrelativistic, low-energy limit of laboratory quantum systems.
While the proposed test in matter wave interferometry experiments turned out to still be five to
six orders of magnitude beyond the state of the art~\cite{Giulini:2011}, tests in optomechanical
set-ups are at the edge of feasible observations~\cite{Yang:2013,Grossardt:2016}.

Here, the underlying physics behind these proposed tests is reviewed. In section~\ref{sec2}
the nonrelativistic limit of Eq.~\eqref{eqn:sce} to the so-called Schrödinger-Newton (SN) equation
is discussed, followed by a review of how one obtains an equation for the centre of mass of
a many-particle system in section~\ref{sec3}. Effects of this modification of the Schrödinger equation
in a harmonic oscillator are analysed in section~\ref{sec4} with a discussion of the possibilities
for testing them in section~\ref{sec5}. Some open questions are addressed in the conclusion
section~\ref{sec6}.

\section{Semi-classical gravity in the nonrelativistic limit}\label{sec2}
Starting with the \emph{classical} Einstein equations for a \emph{classical} Klein-Gordon (or Dirac)
field, one can show that the gravitational self-coupling is promoted to the Schrödinger equation
(as a differential equation for the evolution of a classical field)
obtained in the nonrelativistic ($c \to \infty$) limit~\cite{Giulini:2012}, where it yields
the nonlinear one-particle SN equation:
\begin{equation}
\label{eqn:sn}
\mathrm{i} \hbar \frac{\partial}{\partial t} \psi(t,\vec r) = \left( -\frac{\hbar^2}{2 m} \nabla^2
- G m^2 \int \D^3 \vec r' \, \frac{\abs{\psi(t,\vec r')}^2}{\abs{\vec r - \vec r'}}
\right) \psi(t,\vec r) \,.
\end{equation}

For a \emph{quantum} field, coupling to gravity according to Eq.~\eqref{eqn:sce}, one can proceed
in a similar fashion~\cite{Bahrami:2014}.
In the nonrelativistic, weak coupling limit, Einstein's equations yield
the Poisson equation $\nabla^2 V = 4 \pi G\,m \, \bra{\Psi} \hat{\psi}^\dagger \hat{\psi} \ket{\Psi}$,
while the field equations on curved space-time limit to the Fock space Schrödinger equation including
the solution $V$ of this Poisson equation as a nonlinear potential. The sectors belonging to different
particle numbers separate in the nonrelativistic limit.
In conclusion, one ends up with the $N$-particle equation~\cite{Diosi:1984}
\begin{subequations}\label{eqn:n-particle-sn}\begin{align}
\mathrm{i} \hbar \frac{\partial}{\partial t} \Psi(t,\vec r_1,\cdots,\vec r_N)
&= \biggl(-\sum_{i=1}^N\frac{\hbar^2}{2m_i}\nabla^2_i + V_\text{ext} +  V_g[\Psi] \biggr)\Psi(t;\vec r_1,\cdots \vec r_N) \\
V_g[\Psi](t,\vec r_1,\cdots,\vec r_N) &= -G\sum_{i=1}^N\sum_{j=1}^N m_i m_j \int \mathrm{d}^3 \vec r_1' \cdots \mathrm{d}^3 \vec r_N'
 \frac{\abs{\Psi(t;\vec r'_1,\cdots,\vec r'_N)}^2}{\abs{\vec r_i - \vec r'_j}} \,,\label{eqn:vg-mp}
\end{align}\end{subequations}
where $V_\text{ext}$ stands for any additional non-gravitational potentials.
Eq.~\eqref{eqn:n-particle-sn} contains both the mutual gravitational interactions between particles
and the gravitational self-interaction of each particle. In the one-particle case, $N=1$, one obtains
Eq.~\eqref{eqn:sn}.

\section{The Schrödinger-Newton equation for the centre-of-mass motion}\label{sec3}
Contrary to linear interaction potentials, the wave-function dependent potential~\eqref{eqn:vg-mp}
does not separate into centre of mass and relative coordinates exactly. Note, however, that the
electromagnetic forces which determine the shape of a complex quantum system like a large molecule
are considerably stronger than the gravitational forces. One can, therefore, make use of an
approximation similar to the Born--Oppenheimer approximation in atom physics~\cite{Giulini:2014}.
The characteristic time scale of the relative motion, dominated by the electromagnetic interactions,
is much shorter than that of the centre of mass motion, which is only affected by self-gravitational
forces. The potential~\eqref{eqn:vg-mp} can then be averaged over the relative degrees of freedom,
resulting in a potential that only depends on the centre of mass coordinate, and in the following
centre of mass SN equation:
\begin{subequations}\label{eqn:cm-sn}\begin{align}
\mathrm{i} \hbar \frac{\partial}{\partial t} \psi(t,\mathbf{r})
&= \left( -\frac{\hbar^2}{2\,M}\nabla^2 + V_\text{ext} +  V_g[\psi] \right)\,\psi(t,\mathbf{r}) \\
V_g[\psi](t,\mathbf{r}) &= -G \, \int \D^3 \vec r' \, \abs{\psi(t,\mathbf{r'})}^2 \, I_{\rho_c}(\mathbf{r}
- \mathbf{r'})\label{eqn:vg}\\
I_{\rho_c}(\vec d) &= \int \mathrm{d}^3 \vec u \, \mathrm{d}^3 \vec v \, \frac{\rho_c(\vec u) \, \rho_c(\vec v - \vec d)}{\abs{\vec u - \vec v}} \,,
\label{eqn:grav-self-interaction}
\end{align}
where $M$ is the total mass, $\psi$ the centre of mass wave-function, and $\rho_c$ is the (averaged)
mass distribution of the quantum system, defined from the relative wave-function $\chi$ as
\begin{equation}
\rho_c(\vec r)
=\sum_{i=1}^{N-1} m_i \int \mathrm{d}^3 \vec r_1 \cdots \mathrm{d}^3 \vec r_{i-1} \, \mathrm{d}^3 \vec r_{i+1} \cdots \mathrm{d}^3 \vec r_{N-1} \abs{\chi(\vec r_1,\dots,\vec r_{i-1},\vec r,\vec r_{i+1},\dots,\vec r_{N-1})}^2\,.
\end{equation}\end{subequations}

Now, if the size of the considered system (e.\,g. molecule) is small compared to the extent of its centre of
mass wave-function, $\rho_c$ can be approximated by a delta distribution, and Eq.~\eqref{eqn:cm-sn}
is approximately equivalent to the one-particle equation~\eqref{eqn:sn}.

If the width of the wave-function becomes comparable to the size of the system then its structure,
represented by the function $I_{\rho_c}$, must be taken into account. For a homogeneous spherical
particle of radius $R$, for instance, one finds
\begin{equation}\label{eqn:I-sphere}
I_{\rho_c}^\text{sphere}(d)=\frac{M^2}{R}\times
\begin{cases}
\frac{6}{5}-2\left(\frac{d}{2R}\right)^2+\frac{3}{2}\left(\frac{d}{2R}\right)^3-\frac{1}{5}\left(\frac{d}{2R}\right)^5
&\text{for}\ d\leq 2R\,,\\
\frac{R}{d}
&\text{for}\ d\geq 2R\,.
\end{cases}
\end{equation}
A more realistic model of a molecule also takes into account the crystalline structure of the atoms.
A spherical particle of radius $R$ with atoms localised with a Gaussian distribution of width $\sigma$
is well approximated by the function~\cite{Grossardt:2016,Grossardt:2016a}
\begin{equation}\label{eqn:I-cryst}
I_{\rho_c}^\text{cryst}(d) \approx \frac{6\,M^2}{5\,R} + \frac{M\,m_\text{atom}}{d}\,\mathrm{erf}\left( \frac{d}{\sqrt{2}\,\sigma} \right)\,,
\end{equation}
as long as the wave-function has an extent small compared to the radius $R$ of the whole particle.
The first term stems from the mutual gravitational attraction of the atoms, while the second term
describes the accumulated self-gravitation forces of each atom with its own marginal wave-function.

In the narrow wave-function regime, where the extent of the wave-function is small also compared to
the localisation length $\sigma$ of the atoms, the error function can be expanded around $d = 0$,
yielding the quadratic approximation
\begin{equation}\label{eqn:I-narrow}
I_{\rho_c}^\text{narrow}(d) \approx \frac{6\,M^2}{5\,R}
+ \frac{2\,M\,m_\text{atom}}{\sqrt{2\pi}\sigma}\,\left( 1 - \frac{d^2}{6\,\sigma^2} \right)\,.
\end{equation}

\section{Effects on harmonically trapped quantum systems}\label{sec4}
\begin{figure}
\centering
\includegraphics[scale=0.7]{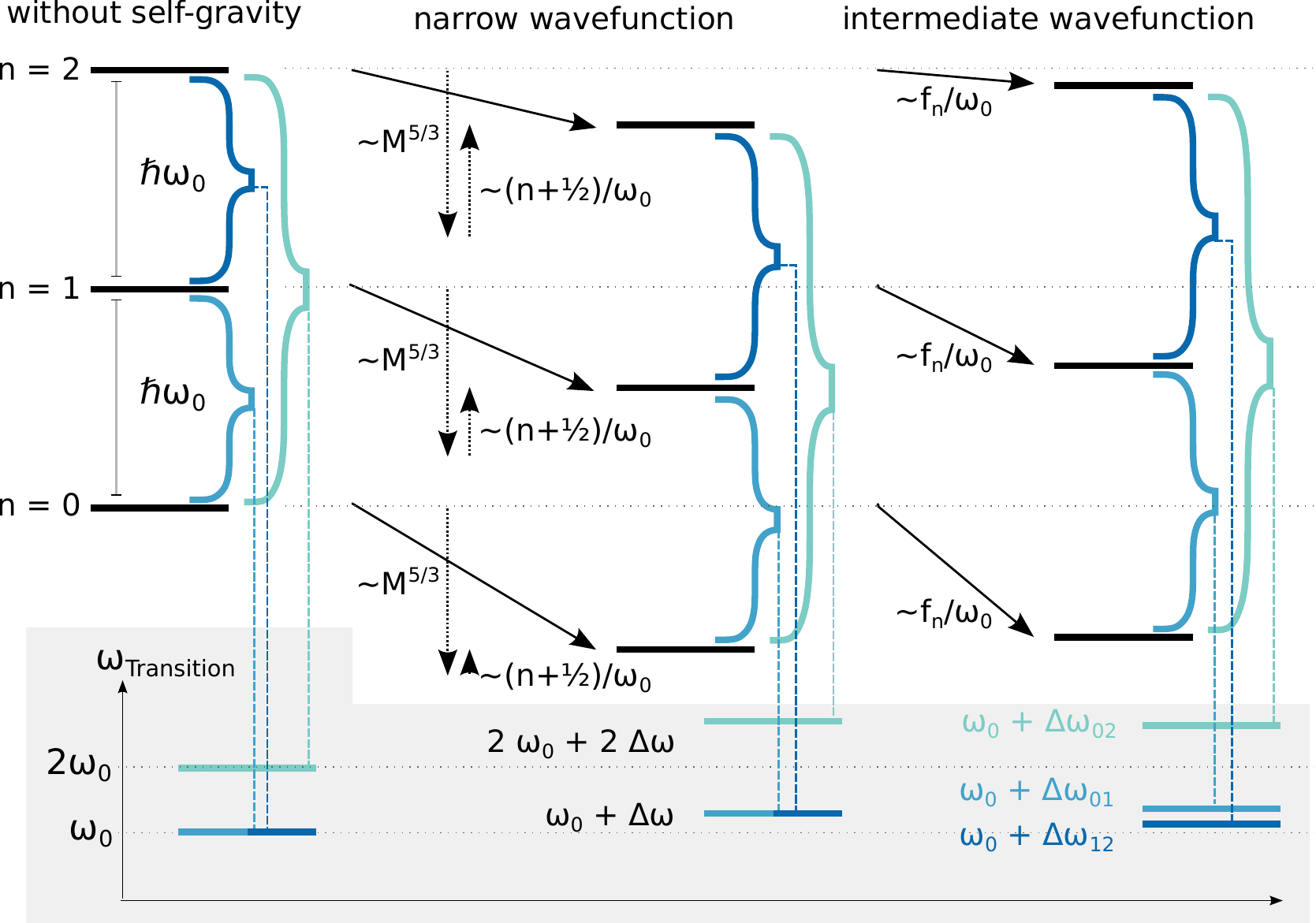}
\caption{Schematic picture of the self-gravitational effect on the harmonic oscillator spectrum in the narrow and intermediate wave-function regime~\cite{Grossardt:2016}.\label{fig:schematic}}
\end{figure}

We assume that the potential $V_\text{ext}$ is such that it strongly confines the wave-function in
the $y$- and $z$-direction. Then the SN equation can be shown to become effectively
one-dimensional~\cite{Grossardt:2016a}. If we assume a quadratic potential in the $x$-direction,
$V_\text{ext} = M\,\omega_0^2\,x^2/2$, then for the narrow wave-function regime, using
Eq.~\eqref{eqn:I-narrow}, one obtains the Hamiltonian
\begin{align}
H^\text{narrow}
&= -\frac{\hbar^2}{2\,M}\nabla^2 + \frac{M}{2}\omega_0^2\,x^2
- \frac{6\,G\,M^2}{5\,R}
- \frac{2\,G\,M\,m_\text{atom}}{\sqrt{2\pi}\sigma}\,\left( 1
- \frac{x^2 -2\,x\,\langle x \rangle+\langle x^2 \rangle}{6\,\sigma^2} \right) \nonumber\\
&= -\frac{\hbar^2}{2\,M}\nabla^2
+ \frac{M}{2}\omega_\text{SN}^2 \, \langle x^2 \rangle
- M\,\omega_\text{SN}^2 \,x\,\langle x \rangle
+ \frac{M}{2}\left(\omega_0^2 + \omega_\text{SN}^2\right)x^2
+ \text{const.} \,,
\intertext{with}
\omega_\text{SN} &= \sqrt{\sqrt{\frac{2}{\pi}}\frac{G\,m_\text{atom}}{3\,\sigma^3}}\,.
\label{eqn:wsn}
\end{align}

For a wave-function comparable to the atom localisation scale, the Hamiltonian becomes a much more
complicated nonlinear functional. However, for the sake of determining the distortion of the energy 
spectrum one can simply use first order perturbation theory. The potential~\eqref{eqn:vg} evaluated
for the unperturbed eigenstates $\psi_n^{(0)}$ is then treated as a perturbation, yielding the
energy shift
\begin{equation}\label{eqn:energy-correction}
\Delta E_n = \bra{\psi^{(0)}_n} V_g[\psi^{(0)}_n](\mathbf{r}) \ket{\psi^{(0)}_n}
= -\frac{G\,\hbar\,m_\text{atom}}{4\,\sigma^3\,\omega_0}\,f_n(\alpha)
\,,\quad\quad\text{with}\quad\quad
\alpha = 2\sigma\,\sqrt{\frac{M\omega_0}{\hbar}} \,.
\end{equation}
In the case of the narrow wave-function, the function $f_n$ depends on the quantum number $n$
\emph{linearly}, implying that the energy gap between eigenstates changes
to $\Delta n\,\hbar\omega_0\,(1+\omega_\text{SN}^2/\omega_0^2)$.
Nonetheless, the transition energy only depends on the difference $\Delta n = (n_1 - n_2)$, not on
$n_1$ and $n_2$ explicitly.

\begin{figure}
\centering
\includegraphics[width=16pc]{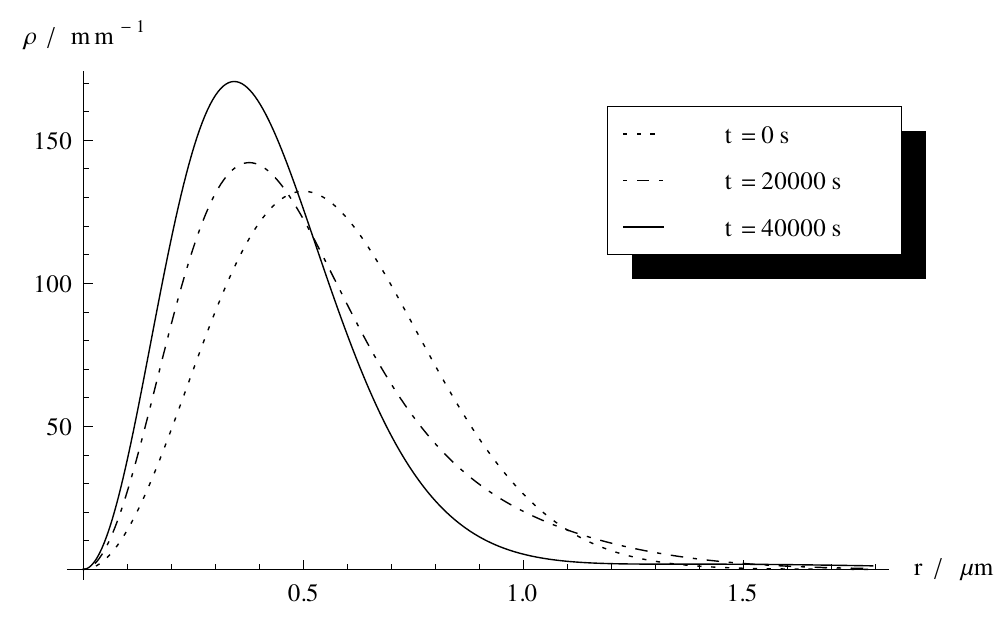}\hspace{2pc}%
\includegraphics[width=16pc]{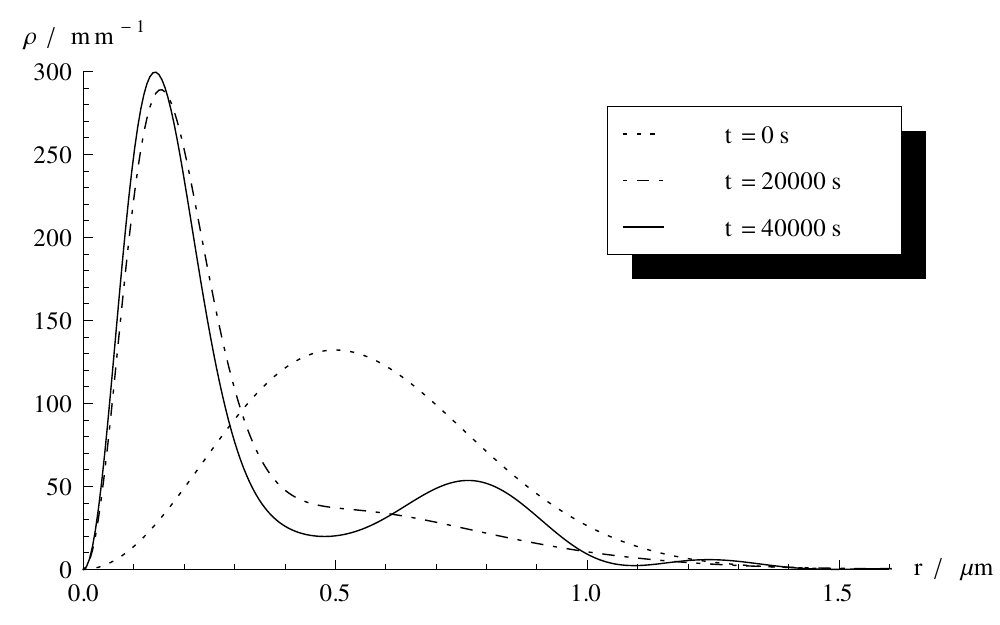}
\caption{Inhibitions of the free dispersion of a wave packet. Plotted is the radial probability 
density $\rho = 4\pi\,r^2\,\abs{\psi}^2$ for a mass of $7\times10^9~\atomicmass$ 
(left) and $10^{10}~\atomicmass$ (right) at different times.\label{fig:dispersion}}
\end{figure}

This changes in the regime of wider wave-functions, where $f_n$ becomes nonlinear in
$n$~\cite{Grossardt:2016,Grossardt:2016a}:
\begin{subequations}\begin{align}
f_n(\alpha) &\approx \text{const.} + \alpha^3\,\sqrt{\frac{2}{\pi}} \, \int_0^\infty \D \zeta \,
\exp \left(-\frac{\alpha^2\,\zeta^2}{2}\right) \, P_n(\alpha\,\zeta) \,
\left(\frac{\mathrm{erf}\left(\sqrt{2}\,\zeta\right)}{2\,\zeta} - \sqrt{\frac{2}{\pi}}\right) \,,\\
P_n(z) &= \frac{\mathrm{e}^{-z^2/2}}{\sqrt{2\pi}\,(2^n \, n!)^2} \,
 \int_{-\infty}^\infty \D \xi \, \mathrm{e}^{-2\xi^2} \,H_n\left( \xi \right)^2
\left( \mathrm{e}^{2\,z\,\xi} \, H_n\left( \xi - z \right)^2
+ \mathrm{e}^{-2\,z\,\xi} \, H_n\left( \xi + z \right)^2 \right) \,,
\end{align}\end{subequations}
where $H_n$ are the Hermite polynomials. Now the frequency gap between adjacent energy levels
depends on $n_1$, $n_2$ explicitly, yielding a fine structure for the transition frequencies as depicted
in Fig.~\ref{fig:schematic}.

Instead of the energy spectrum, one can study the effect of the self-gravitational potential on
the dynamics of a quantum harmonic oscillator. For a squeezed coherent (i.\,e. Gaussian) state,
the dynamics can be reduced to a set of two differential equations for the first and second moments:
\begin{subequations}\begin{align}
0 &= \frac{\mathrm{d}^2}{\mathrm{d} t^2}\erw{x} + \omega_0^2 \erw{x} \\
0 &= \frac{\mathrm{d}^3}{\mathrm{d} t^3}\erw{(x-\erw{x})^2} + 4\,\left( \omega_0^2
+ \omega_\text{SN}^2[\psi]\right)\,\frac{\mathrm{d}}{\mathrm{d} t}\erw{(x-\erw{x})^2} \,.
\end{align}\end{subequations}
The first equation describes the usual oscillation of the centre of the wave-function with the
unperturbed trap frequency $\omega_0$. The second equation describes the internal oscillation,
for which the frequency is shifted by a factor of $\sqrt{1 + \omega_\text{SN}^2/\omega_0^2}$.

In the general case of a wider wave-function this frequency shift depends on the wave-function. However,
it is strongest for a narrow wave-function, where $\omega_\text{SN}$ takes the value~\eqref{eqn:wsn}.
For a wider wave-function, the self-gravitational effect only becomes less significant.
The effect in the narrow regime was discussed by Yang et~al.~\cite{Yang:2013}.

\section{Experimental tests}\label{sec5}

First ideas to test the SN equation experimentally~\cite{Carlip:2008} were based on the loss of
interference in matter-wave interferometry experiments~\cite{Arndt:2014}. Current experiments
reach masses of up to $10^4~\atomicmass$ and are clearly in the wide wave-function regime
were Eq.~\eqref{eqn:sn} is valid. Numerical and analytical estimates~\cite{Giulini:2011} yield a
required mass around $10^{10}~\atomicmass$ for a micrometre sized wave packet to show
significant inhibitions of the free dispersion, see Fig.~\ref{fig:dispersion} for example
plots of the evolution of a Gaussian wave packet. The required mass values seem infeasible on
Earth, although they might be achievable in space experiments~\cite{Kaltenbaek:2016}.
Note also that for the required mass scales, the effects of the internal structure of the particles
are no longer negligible and, therefore, Eq.~\eqref{eqn:sn} is only of limited use for good
predictions. However, approximation schemes exist~\cite{Grossardt:2016b,Colin:2016} which allow
for a simple numerical calculation of the magnitude of the deviation between the nonlinear dynamics
according to the SN equation and the ordinary, linear Schrödinger equation.

\begin{figure}
\centering
\includegraphics[width=20pc]{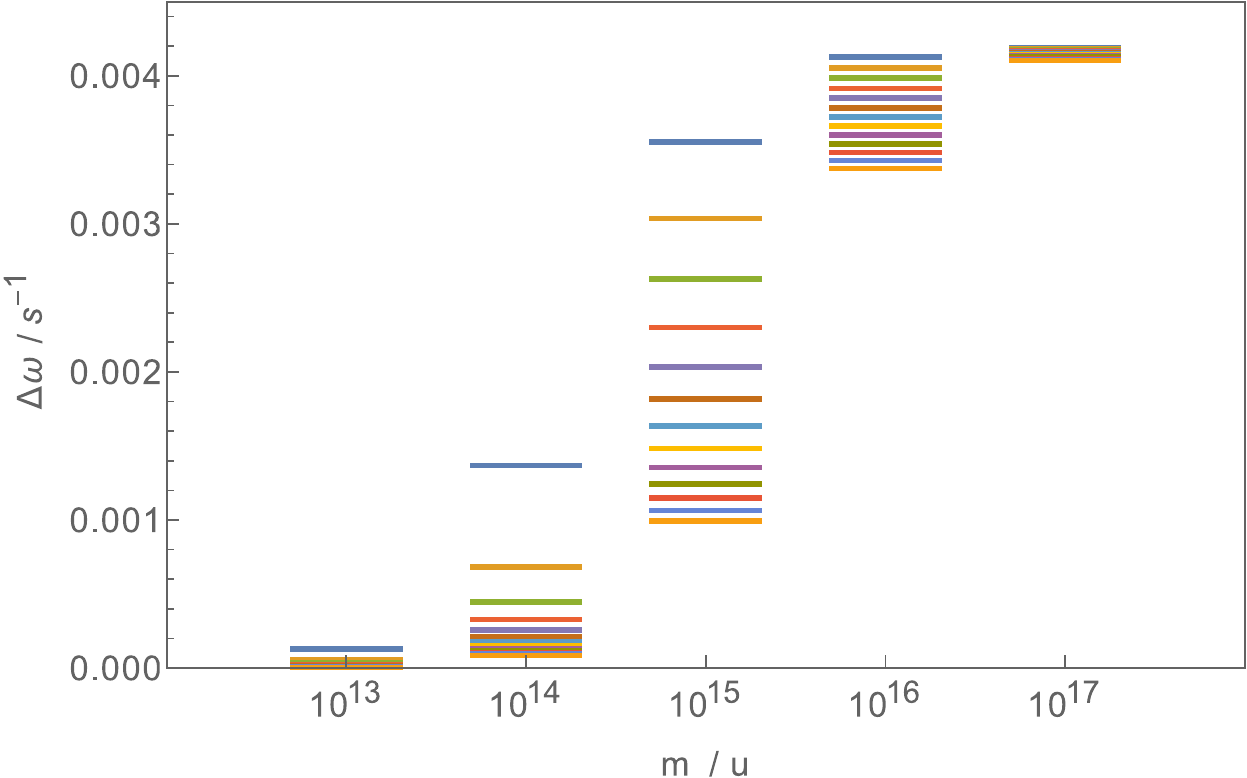}\hspace{2pc}%
\begin{minipage}[b]{12pc}\caption{Resulting frequency spectrum of the transition frequency
$\Delta \omega$ between adjacent eigenstates ($\Delta n = 1$) for osmium at $T = 100\,\milli\kelvin$,
$\omega_0 = 2\pi\times10\,\second^{-1}$.\label{fig:plot_spectrum}}
\end{minipage}
\end{figure}

As far as ground based experiments are concerned, the two effects discussed here for particles
trapped in quadratic potentials seem the most promising routes towards an experimental test.
A proposal for a test of the spectral effect was presented in Ref.~\cite{Grossardt:2016}.
It consists of a levitated osmium disc in a Paul trap which is cooled to a temperature below
100~\milli\kelvin. At this temperature osmium becomes superconducting, avoiding decoherence
due to blackbody radiation. The suggested experimental parameters are a particle mass of
$10^{14}~\atomicmass$ trapped at a frequency of 10~\hertz. At these values, a frequency resolution
of 0.1~\milli\hertz\ would be sufficient for a detection of the spectral fine structure due to
the SN equation. Fig.~\ref{fig:plot_spectrum} shows a quantitative plot of the expected frequency
spectrum for these values. At low masses, self-gravity becomes negligible. At high masses all spectral
lines are shifted equally. The intermediate regime, where a significant splitting appears,
spans about 3 orders of magnitude in mass.

The crucial parameter for the choice of material is the Debye--Waller factor, which determines the
localisation $\sigma$ of the atoms within the crystal. It takes an ideal value for
osmium, even compared to materials of larger atomic mass such as gold; see Tab.~\ref{tab:material}
for a comparison of some materials.
Levitation of a micrometre particle in a Paul trap has already been demonstrated
experimentally~\cite{Wuerker:1959}, as have the low frequencies required~\cite{Gerlich:2003},
and the control of decoherence~\cite{Poulsen:2012}. Cooling to low energy states has not been
demonstrated for the required masses, but the technological tools exist.

The dynamical effects~\cite{Yang:2013} are somewhat disadvantageous to observe: they require both preparation of
a squeezed state and a larger mass in order to reach the narrow wave-function regime.
On the other hand, the required set-up is closer to existing ones, such as cooled mirrors (e.\,g.
LIGO or LISA pathfinder). The fact that all of these systems are based on silicon rather than
osmium, however, requires another two orders of magnitude better resolution compared to the
same system based on osmium (cf. Tab.~\ref{tab:material}). At 10~\hertz\ trap frequency, a quality
factor better than $10^6$ is required~\cite{Yang:2013}.
The feasibility for a test in several systems has been examined in Ref.~\cite{Gan:2016}.

\begin{table}
\caption{Relevant experimental parameters for different materials~\cite{Grossardt:2016}.}\label{tab:material}
\begin{center}
\begin{tabular}{lrrrr}
\br
Material & $m_\text{atom}$ / u & $\rho$ / g\,cm\textsuperscript{-3} & $\sigma$ / pm
& $\omega_\text{SN}^2 / \omega_0^2$ \\
\hline
Silicon & 28.086 & 2.329 & 6.96 & 0.00246 \\
Tungsten & 183.84 & 19.30 & 3.48 & 0.128 \\
Osmium & 190.23 & 22.57 & 2.77 & 0.264 \\
Gold & 196.97 & 19.32 & 4.66 & 0.0574 \\
\br
\end{tabular}
\end{center}
\end{table}

\section{Conclusion and open questions}\label{sec6}
From the above discussion we conclude that, although a realisation is still challenging,
the SN equation seems testable with existing technology in the nearer future.
What would we learn from such a test?

Obviously, if the predicted deviations from linear quantum mechanics would be observed,
this would result in a complete paradigm shift of our view on both gravity and the foundations
of quantum mechanics. If, on the other hand, standard quantum mechanics is confirmed and the
SN equation ruled out by experiments (likely the outcome expected by a large majority of physicists)
this would rule out the semi-classical Einstein equations~\eqref{eqn:sce} as a fundamental model.
It would, however, not be conclusive evidence for the quantisation of gravity, since other types of
coupling of quantum matter to a classical space-time could be possible, which do not yield the
SN equation as a nonrelativistic limit. In fact, with stochastic
gravity~\cite{Hu:2008,Anastopoulos:2014a} there is an example for such a model.

As the SN equation was first discussed by Diósi~\cite{Diosi:1984} and Penrose~\cite{Penrose:1998}, 
it is often considered in context of the gravitational collapse model named after these
two~\cite{Diosi:1987,Diosi:1989,Penrose:1996,Penrose:1998,Penrose:2014}.
Although not directly related to it, one can discuss whether the SN equation could play a role
in the collapse of the wave-function. On the one hand, the classicality of the gravitational
force provides exactly the non-unitary ingredient needed for an objective collapse. On the other hand,
the SN alone does not explain the emergence of classicality, since the SN ``collapse'' is completely
deterministic. It does not recover the Born probability rule~\cite{Bahrami:2014}.

Probably the biggest concern against a fundamental validity of the SN equation is that measurements
in the usual instantaneous collapse prescription allow for faster-than-light signalling when combined
with a nonlinear dynamics~\cite{Bahrami:2014}. It might be that a consistent description
which makes use of semi-classical gravity as an explanation of the wave-function collapse can
circumvent this problem. This description, however, has yet to be found.

\ack
The advances reviewed here resulted from collaborations with M. Bahrami, A. Bassi, J. Bateman, 
S. Donadi, D. Giulini, and H. Ulbricht. I gratefully acknowledge funding from the German Research
Foundation (DFG) and the Italian National Institute for Nuclear Physics (INFN).

\section*{References}
\providecommand{\newblock}{}


\begin{thebibliography}{10}
\expandafter\ifx\csname url\endcsname\relax
  \def\url#1{{\tt #1}}\fi
\expandafter\ifx\csname urlprefix\endcsname\relax\def\urlprefix{URL }\fi
\providecommand{\eprint}[2][]{\url{#2}}

\bibitem{Moller:1962}
Møller C 1962 {\em Colloques Internationaux CNRS\/} vol~91 ed Lichnerowicz A
  and Tonnelat M~A (CNRS, Paris)

\bibitem{Rosenfeld:1963}
Rosenfeld L 1963 {\em Nucl. Phys.\/} {\bf 40} 353--356

\bibitem{Page:1981}
Page D~N and Geilker C~D 1981 {\em Phys. Rev. Lett.\/} {\bf 47} 979--982

\bibitem{Bassi:2013}
Bassi A, Lochan K, Satin S, Singh T~P and Ulbricht H 2013 {\em Rev. Mod.
  Phys.\/} {\bf 85} 471--527 (\textit{Preprint} \eprint{1204.4325})

\bibitem{Eppley:1977}
Eppley K and Hannah E 1977 {\em Found. Phys.\/} {\bf 7} 51--68

\bibitem{Mattingly:2005}
Mattingly J 2005 {\em Einstein Studies Volume 11. The Universe of General
  Relativity\/} Einstein Studies ed Kox A~J and Eisenstaedt J (Boston:
  Birkh\"auser) chap~17, pp 327--338

\bibitem{Kiefer:2007}
Kiefer C 2007 {\em Quantum Gravity\/} 2nd ed ({\em International Series of
  Monographs on Physics\/} vol 124) (Oxford: Clarendon Press)

\bibitem{Albers:2008}
Albers M, Kiefer C and Reginatto M 2008 {\em Phys. Rev. D\/} {\bf 78} 064051
  (\textit{Preprint} \eprint{0802.1978})

\bibitem{Carlip:2008}
Carlip S 2008 {\em Class. Quantum Grav.\/} {\bf 25} 154010 (\textit{Preprint}
  \eprint{0803.3456})

\bibitem{Giulini:2011}
Giulini D and Großardt A 2011 {\em Class. Quantum Grav.\/} {\bf 28} 195026
  (\textit{Preprint} \eprint{1105.1921})

\bibitem{Yang:2013}
Yang H, Miao H, Lee D~S, Helou B and Chen Y 2013 {\em Phys. Rev. Lett.\/} {\bf
  110} 170401 (\textit{Preprint} \eprint{1210.0457})

\bibitem{Grossardt:2016}
Großardt A, Bateman J, Ulbricht H and Bassi A 2016 {\em Phys. Rev. D\/} {\bf
  93} 096003 (\textit{Preprint} \eprint{1510.01696})

\bibitem{Giulini:2012}
Giulini D and Großardt A 2012 {\em Class. Quantum Grav.\/} {\bf 29} 215010
  (\textit{Preprint} \eprint{1206.4250})

\bibitem{Bahrami:2014}
Bahrami M, Großardt A, Donadi S and Bassi A 2014 {\em New J. Phys.\/} {\bf 16}
  115007 (\textit{Preprint} \eprint{1407.4370})

\bibitem{Diosi:1984}
Diósi L 1984 {\em Phys. Lett. A\/} {\bf 105} 199--202

\bibitem{Giulini:2014}
Giulini D and Großardt A 2014 {\em New J. Phys.\/} {\bf 16} 075005
  (\textit{Preprint} \eprint{1404.0624})

\bibitem{Grossardt:2016a}
Großardt A, Bateman J, Ulbricht H and Bassi A 2016 {\em Sci. Rep.\/} {\bf 6}
  30840 (\textit{Preprint} \eprint{1510.01262})

\bibitem{Arndt:2014}
Arndt M and Hornberger K 2014 {\em Nat. Phys.\/} {\bf 10} 271--277

\bibitem{Kaltenbaek:2016}
Kaltenbaek R, Arndt M, Aspelmeyer M, Barker P~F, Bassi A, Bateman J, Bongs K,
  Bose S, Braxmaier C, Brukner C, Christophe B, Chwalla M, Cohadon P~F, Cruise
  A~M, Curceanu C, Dholakia K, Döringshoff K, Ertmer W, Gieseler J, Gürlebeck
  N, Hechenblaikner G, Heidmann A, Herrmann S, Hossenfelder S, Johann U, Kiesel
  N, Kim M, Lämmerzahl C, Lambrecht A, Mazilu M, Milburn G~J, Müller H,
  Novotny L, Paternostro M, Peters A, Pikovski I, Pilan-Zanoni A, Rasel E~M,
  Reynaud S, Riedel C~J, Rodrigues M, Rondin L, Roura A, Schleich W~P,
  Schmiedmayer J, Schuldt T, Schwab K~C, Tajmar M, Tino G~M, Ulbricht H, Ursin
  R and Vedral V 2016 {\em EPJ Quant. Tech.\/} {\bf 3} 5 (\textit{Preprint}
  \eprint{1503.02640})

\bibitem{Grossardt:2016b}
Großardt A 2016 {\em Phys. Rev. A\/} {\bf 94} 022101 (\textit{Preprint}
  \eprint{1503.02622})

\bibitem{Colin:2016}
Colin S, Durt T and Willox R 2016 {\em Phys. Rev. A\/} {\bf 93} 062102
  (\textit{Preprint} \eprint{1402.5653})

\bibitem{Wuerker:1959}
Wuerker R~F, Shelton H and Langmuir R~V 1959 {\em J. Appl. Phys.\/} {\bf 30}
  342--349

\bibitem{Gerlich:2003}
Gerlich D 2003 {\em Hyperfine Interactions\/} {\bf 146-147} 293--306 ISSN
  0304-3843

\bibitem{Poulsen:2012}
Poulsen G, Miroshnychenko Y and Drewsen M 2012 {\em Phys. Rev. A\/} {\bf 86}(5)
  051402

\bibitem{Gan:2016}
Gan C, Savage C and Scully S 2016 {\em Phys. Rev. D\/} {\bf 93} 124049

\bibitem{Hu:2008}
Hu B~L and Verdaguer E 2008 {\em Living Rev. Relativity\/} {\bf 11} 3

\bibitem{Anastopoulos:2014a}
Anastopoulos C and Hu B~L 2014 {\em New J. Phys.\/} {\bf 16} 085007
  arXiv:1403.4921 (\textit{Preprint} \eprint{1403.4921})

\bibitem{Penrose:1998}
Penrose R 1998 {\em Phil. Trans. R. Soc. A\/} {\bf 356} 1927--1939

\bibitem{Diosi:1987}
Diósi L and Lukács B 1987 {\em Ann. Phys. (Berlin)\/} {\bf 499} 488--492

\bibitem{Diosi:1989}
Diósi L 1989 {\em Phys. Rev. A\/} {\bf 40} 1165--1174

\bibitem{Penrose:1996}
Penrose R 1996 {\em Gen. Relativ. Gravit.\/} {\bf 28} 581--600

\bibitem{Penrose:2014}
Penrose R 2014 {\em Found. Phys.\/} {\bf 44} 557--575

\end{thebibliography}
\end{document}